# A System for Accurate Tracking and Video Recordings of Rodent Eye Movements using Convolutional Neural Networks for Biomedical Image Segmentation

Isha Puri          David D. Cox
Center for Brain Science, Harvard University, Cambridge, MA

*Abstract*— **Research in neuroscience and vision science relies heavily on careful measurements of animal subject's gaze direction. Rodents are the most widely studied animal subjects for such research because of their economic advantage and hardiness. Recently, video based eye trackers that use image processing techniques have become a popular option for gaze tracking because they are easy to use and are completely non-invasive. Although significant progress has been made in improving the accuracy and robustness of eye tracking algorithms, unfortunately, almost all of the techniques have focused on human eyes, which does not account for the unique characteristics of the rodent eye images, e.g., variability in eye parameters, abundance of surrounding hair, and their small size. To overcome these unique challenges, this work presents a flexible, robust, and highly accurate model for pupil and corneal reflection identification in rodent gaze determination that can be incrementally trained to account for variability in eye parameters encountered in the field. To the best of our knowledge, this is the first paper that demonstrates a highly accurate and practical biomedical image segmentation based convolutional neural network architecture for pupil and corneal reflection identification in eye images. This new method, in conjunction with our automated infrared video-based eye recording system, offers the state of the art technology in eye tracking for neuroscience and vision science research for rodents.**

## I. INRODUCTION

Eye tracking methods are deployed in a variety of areas, including neurophysiology, health sciences, and human computer interaction. Due to the difficulties in availability of human subjects and inherent safety concerns, the majority of health science and neurophysiology research relies on the use of rodents to study a variety of behavioral and visual phenomena [1, 4, 5, 16, 17]. Traditionally, different eye tracking methods in neurophysiology research can be classified into two broad categories: invasive and non-invasive methods.

Invasive methods are used in non-human subjects and are known to yield high accuracy results. Some of the more popular invasive methods include scleral search coils [8], and electro-oculography [11]. Scleral search coils involve gluing a coil to the corneal surface and present significant challenges for use in rodents because of need of a surgical procedure and the small size of rodent eyes, as well as weight of the coils on those small eyes interfere with rodent eye movements. Electro-oculography is a technique for measuring the corneo-retinal standing potential that exists between the front and the back of the eye. To measure eye movement, pairs of electrodes are typically placed either above and below the eye or to the left and right of the eye. If the eye moves from center position toward one of the two electrodes, this electrode senses the positive side of the retina and the opposite electrode senses the negative side of the retina. Consequently, a potential difference occurs between the electrodes. Assuming that the resting potential is constant, the recorded potential is a measure of the eye's position. This class of methods also suffer from similar issues as search coils for rodent subjects.

Recently, Non-invasive video based eye trackers [12, 16, 18] that use image processing techniques have become a popular option for gaze tracking because they are easy to use and are completely non-invasive. However, these eye trackers typically require a calibration procedure in which the subject must look at a series of points at known gaze angles. While it is possible to rely on natural sight orientation for calibration in human and some non-human species, rodents do not reliably saccade to visual targets, making this form of calibration impossible. To overcome this problem, we developed a fully automated infrared video eye-tracking method that auto-calibrated by estimating the optical geometry of the mouse cornea relative to the camera [16]. Although the system setup for our method has proven to be very robust, it deployed a complex image processing pipeline that has been relatively sensitive to the high variability in rodent eye parameters like size, color etc. Unfortunately, almost all of the recent progress in eye image processing techniques has been made with human eyes [19], and it does not account for the unique characteristics of the rodent eye images, e.g., variability in eye parameters, abundance of surrounding hair, and their small size.

To overcome these unique challenges, this work presents a flexible, robust, and highly accurate model for pupil and corneal reflection identification in rodent gaze determination that can be incrementally trained to account for variability in eye parameters encountered in the field. To the best of our knowledge, this is the first paper that demonstrates a highly accurate and practical biomedical image segmentation based convolutional neural network architecture for pupil and corneal reflection identification in eye images. This new method, in conjunction with our automated infrared video-based eye recording system, offers the state of the art technology in eye tracking for neuroscience and vision science research for rodents.

## II. RODEN GAZE DETECTION SYSTEM

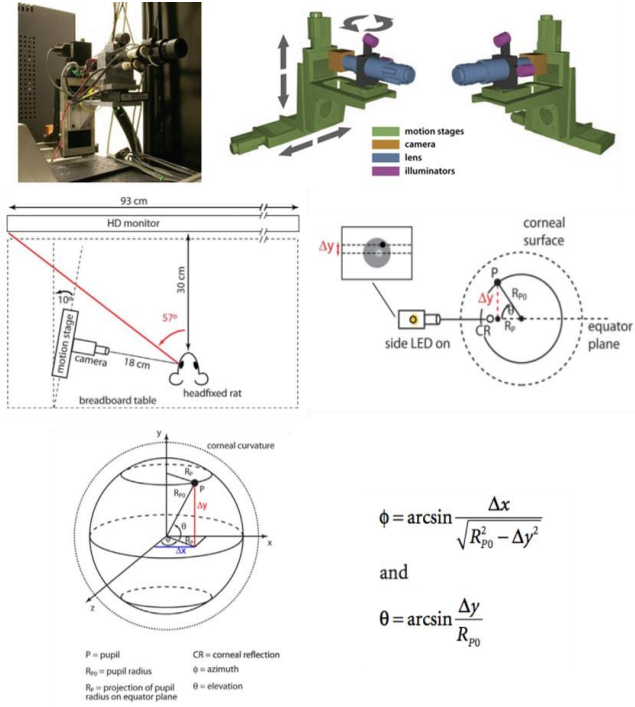

Figure 1: Our tracking and video recording system for Rodent eye movements [16]

In Figure 1, we show our video-based system setup [16] for rodent eye imaging and tracking, where by accurately locating the pupil and corneal reflection on an image of the eye, we can find the gaze angle. Our eye tracking model is based on the fact that because the rodent is head fixed, calculating the gaze angle is an acceptable way to track eye movement. In order to calculate the gaze angle, we need to determine the positions of the pupil and corneal reflection. Our methodology also relies on two main practical assumptions that have both been supported by literature on rodent eye geometry [3, 9]. First, we assume that the pupil rotates about the center of the eyeball. Second, we assume that the corneal curvature of the rodent is almost an exact sphere. These assumptions allow us to rely only on the locations of the corneal reflection and the pupil on the image captured by the camera to find the gaze angle. By calculating the gaze angle for each consecutive frame of a video, we successfully build a highly accurate and robust eye tracking model for the rodent gaze.

In Figure 2, we show the previous, rather complex image processing pipeline [16] that deployed radial geometry transform, rough center determination, Sobel filtering, ellipse fitting, and the Starbust algorithm to identify the pupil and the corneal reflection. This method was relatively sensitive to the high variability in rodent eye parameters. Over years, although our system setup has proven highly practical, the complex eye image processing algorithms and pipeline shown in Figure 2 have suffered from its inability to adapt to the unique characteristics of the rodent eye images, e.g., variability in eye parameters, abundance of surrounding hair, and their small size among over time. Although significant progress has been made in pupil and corneal reflection detection based eye tracking algorithms over last several years, these techniques have mainly focused on human eyes.

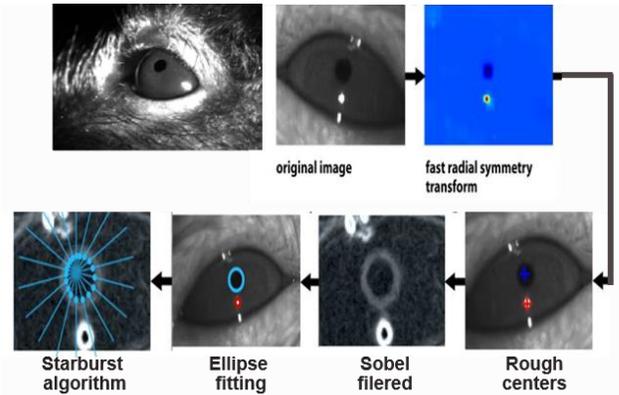

Figure 2: Previous eye image processing pipeline [16]

In the following section, we discuss our new method which deploys Convolutional Neural Networks (CNN) for Biomedical Image Segmentation to overcome these challenges.

## III. PROPOSED METHOD

To overcome the challenges of our problem, this work presents a flexible, robust, and highly accurate model for identification of pupil and corneal reflection. Our new method is a Convolutional Neural Network model called U-Net [14]. As shown in Figure 3, U-Net model consists of a contracting path to capture context in the eye images and a symmetric expanding path that enables highly accurate localization of the pupil and corneal reflection and learns from a small amount of training data. The U-Net is a neural network specialized for image segmentation (Figure 3); it can be trained end-to-end from very few images and outperforms the prior best methods in its task of segmenting a given image, as shown in Figure 4. Key differences between the U-Net and a standard convolutional deep neural network include: a larger number of feature channels, which allow the network to propagate context information to higher resolution layers, and a lack of any fully connected layers, as the network only uses the valid part of each convolution. (See Figure 3). The U-Net algorithm was ideally suited for an eye tracking model for rodents because of its ability to train efficiently on a small number of rather low quality images: the dataset used for this research had only hundreds of training images, and the images were not of extremely high resolution. In addition, U-Net algorithm can segment a complex image into individual features and recognize them with very high accuracy (99%), making them ideally suited for the task of recognizing pupil and corneal reflection features in a complex rodent eye image.

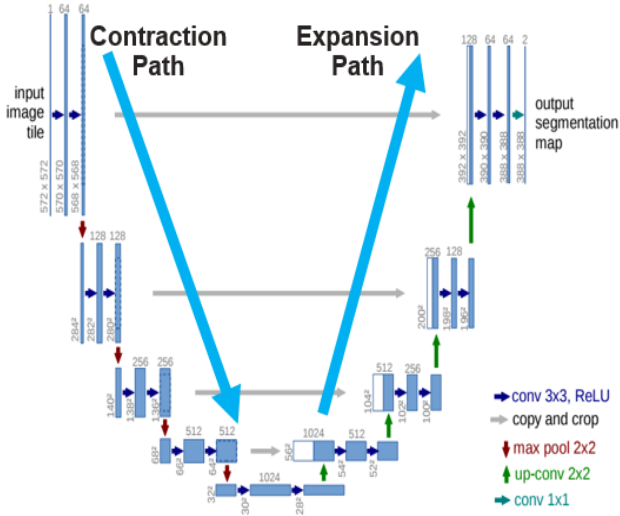

Figure 3: U-Net architecture. The large number of feature channels that compensate for a fully connected neural net layer yield a U-shaped architecture.

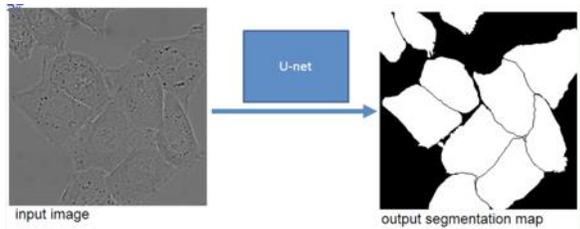

Figure 4: U-Net: Convolutional Neural Network for Biomedical Image segmentation [14]

For our task, we designed and trained two different U-Net models: one to identify the pupil and another to identify the corneal reflection. First we generated the training masks, i.e., ground truth (as shown in Figure 5), for the neural network training for both pupil and corneal reflection detection model using manual labeling for few hundred images (small training data is one of the key constraints in the current task because of the need of incremental training). In general, the corneal reflection identification is harder relative to pupil identification due to the presence of similar features, i.e., presence of many other white spots due to presence of hairs in the region surrounding the eye, as shown in the eye image in Figure 4. We trained this model on a relatively small dataset of images (245 labeled images for pupil and 325 labeled images for corneal reflection) taken from our infrared video eye tracker system (Figure 1) with the corresponding ground truth specified with mask images.

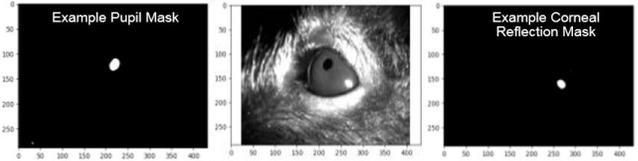

Figure 5: Training Mask generation (ground truth data) for the rodent eye image U-Net CNN Model: Example Pupil Mask (left); Example Corneal Reflection Mask (Right)

IV.　　RESULTS

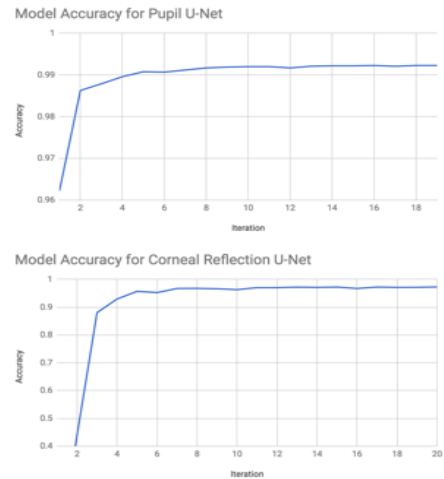

Figure 6: U-Net training curves (Epoch vs Accuracy) for rodent eye tracking models: Pupil U-Net Model (top) with over 99.5% accuracy; Corneal Reflection U-Net Model (bottom) with over 98% accuracy

We tested our new approach extensively and results show that it can **achieve 99.51% accuracy for the Pupil U-Net model and 98.02% accuracy for the corneal reflection U-Net model** *(Figure 6).* As shown in a sample video frame in Figure 7, these two U-Net models working in conjunction can detect pupil (red) and corneal reflection (green) highly accurately. Most importantly, this model can be incrementally trained very quickly within minutes on a single P100 Nvidia GPU computing system, given new data, which allows us to adapt to variability in rodent subjects very quickly, a key metrics for our success. In addition, compared to the previous complex image processing pipeline, this new approach is a single stage model, with its many advantages outlined earlier that help overcome the unique challenges of our task. i.e., adapting perfectly to the unique characteristics of the rodent eye images: variability in eye parameters, abundance of surrounding hair, and their small size. Due to the single stage model, its latency for processing image data is sub 250ms enabling real time eye tracking. With the accurate locations of the pupil and corneal reflection, we can determine the gaze angle using our established system setup in Figure 1. This new method, in conjunction with our automated infrared video-based eye recording system [16], offers the state of the art technology in eye tracking for neuroscience and vision science research for rodents.

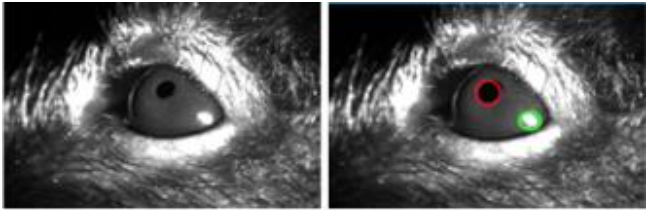

Figure 7: A sample video frame processing result, two U-Net models working in conjunction can detect pupil (red) and corneal reflection (green) highly accurately.

V. CONCLUSIONS

Research in neuroscience and vision science relies heavily on careful measurements of animal subject's gaze direction. Significant progress in the accuracy and robustness of eye tracking algorithms has so far focused mainly on human eyes, and such algorithms are not very sensitive to characteristics of the rodent eye images, e.g., variability in eye parameters, abundance of surrounding hair, and their small size. This work presents a flexible, robust, and highly accurate model for pupil and corneal reflection identification in rodent gaze determination that can easily be incrementally trained to account for variability in eye parameters encountered in rodent eye images. To the best of our knowledge, this is the first paper that demonstrates a highly accurate and practical biomedical image segmentation based convolutional neural network architecture for pupil and corneal reflection identification in eye images. In conjunction with an automated video based eye recording system, offers the state of the art technology in eye tracking for neuroscience and vision science research for rodents.